\documentclass[12pt,preprint]{aastex}

\begin{document}

\newcommand{\xc}{{x^{(2)}}}
\def\lta{\lower2pt\hbox{$\buildrel {\scriptstyle <} 
   \over {\scriptstyle\sim}$}}
\def\gta{\lower2pt\hbox{$\buildrel {\scriptstyle >} 
   \over {\scriptstyle\sim}$}}

\title{A Turbulent Model of Gamma-Ray Burst Variability}

\author{Ramesh Narayan$^1$ \& Pawan Kumar$^2$ \\
$^1$Harvard-Smithsonian Center for Astrophysics, 60 Garden Street, Cambridge, MA 02138
$^2$Astronomy Department, University of Texas, Austin, TX 78712\\
}
\author{Authors}

\begin{abstract}
A popular paradigm to explain the rapid temporal variability observed
in gamma-ray burst (GRB) lightcurves is the internal shock model.  We
propose an alternative model in which the radiating fluid in the GRB
shell is relativistically turbulent with a typical eddy Lorentz factor
$\gamma_t$.  In this model, all pulses in the gamma-ray lightcurve are
produced at roughly the same distance $R$ from the center of the
explosion.  The burst duration is $\sim R/c\Gamma^2$, where $\Gamma$
is the bulk Lorentz factor of the expanding shell, and the duration of
individual pulses in the lightcurve is $\sim R/c\Gamma^2\gamma_t^2$.
The model naturally produces highly variable lightcurves with
$\sim\gamma_t^2$ individual pulses.  Even though the model assumes
highly inhomogeneous conditions, nevertheless the efficiency for
converting jet energy to radiation is high.

\end{abstract}


\section{Introduction}

Our understanding of gamma-ray bursts (GRBs) has improved enormously
in the last 15 years, thanks to observations by dedicated
$\gamma$-ray/X-ray satellites such as Compton-GRO, BeppoSAX, HETE-2
and Swift, and follow-up observations from the ground in optical and
radio (for recent reviews see M\'esz\'aros, 2002; Piran 2005; Woosley
\& Bloom, 2006; Fox \& M\'esz\'aros, 2006; Zhang, 2007).  As a result
of this work, it is now established that at least some long-duration
bursts are produced in the collapse of a massive star (as suggested by
Woosley 1993; Paczynski 1998), accompanied by the ejection of a highly
relativistic jet.  Afterglow observations as well as energy
considerations indicate that the jet is well collimated (Rhoads 1999,
Frail et al. 2001, Panaitescu \& Kumar, 2001).  The presence of a
relativistic jet has also been directly confirmed in the nearby burst
GRB 030329 which exhibited ``superluminal'' motion in its radio
afterglow (Taylor et al. 2004).

GRB lightcurves are known to be highly variable (Meegan et al. 1992),
and this has led to the development of the {\it internal shock model}
(Piran, Shemi \& Narayan 1993; Rees \& Meszaros 1994; Katz
1994). According to this model, the Lorentz factor of the jet varies
with time, and as a result, faster portions of the jet catch up with
slower portions.  In the ensuing collision, a fraction of the kinetic
energy of the jet is converted to thermal energy.  The observed
radiation is then produced via synchrotron and inverse-Compton
processes.

In a seminal paper, Sari and Piran (1997) provided a general argument
why GRB lightcurve variability cannot be explained by simply appealing
to a highly inhomogeneous source.  They showed that the efficiency for
converting jet energy to the observed radiation is extremely poor when
variability arises purely from inhomogeneity; the essence of their
argument is summarized in \S2.  The argument is both powerful and
compelling, and it has led to wide acceptance of the internal shock
model.  A basic feature of the model is that different pulses within
the lightcurve of a GRB are produced in distinct internal
collisions/shocks, generally at different distances from the central
explosion.

There are, however, a number of observations and/or theoretical
considerations that pose difficulties for the internal shock model.
Internal shocks have only a modest efficiency $\sim 1$--10\% for
converting jet energy to the radiation observed in the
$20$\,keV--1\,MeV band (Kumar, 1999; Panaitescu, Spada \& Meszaros,
1999; Lazzati et al. 1999)\footnote{Beloborodov (2000) and Kobayashi
\& Sari (2001) reported a much higher efficiency $\sim100$\%.
However, their estimates were based purely on the kinematics of
colliding shells, where shells of Lorentz factor $\gamma\sim1$
collided with high $\gamma$ shells.  They did not take into
consideration the efficiency of the emergent radiation in the commonly
observed 20\,keV--1\,MeV band.}.  Even a $\sim$10\% radiative
efficiency is low compared to the burst efficiency implied by
measurements of the jet kinetic energy through modeling GRB afterglow
lightcurves (Panaitescu \& Kumar, 2002).

Another difficulty with the internal shock model is the large distance
from the central explosion that one estimates ($R\ \gta\ 10^{16}$cm)
for the $\gamma$-ray-producing region in a number of GRBs (Kumar et
al. 2007).  This distance is significantly larger than what one
expects in the internal shock model.  Moreover, the estimated distance
is within a factor of a few of the deceleration radius where the jet
begins to interact with the external medium.  This coincidence between
two unrelated radii is unexpected.

These difficulties, along with the problem of avoiding excessive
baryon loading, motivate us to consider an alternative to the internal
shock model.  We show in \S3 that the argument of Sari \& Piran (1997)
against an inhomogeneous source can be successfully overcome if we
consider source inhomogeneities that move with {\it relativistic
velocities}.  In \S4, we calculate a model lightcurve using this
relativistic model and demonstrate that it is consistent with
observations of GRBs.  We conclude in \S5 with a discussion.

\section{Internal Shock Model}

We consider an idealized model of a GRB in which a spherical shell,
located at radius $R$ with respect to the center of the explosion,
expands ultra-relativistically outward with a bulk Lorentz factor
$\Gamma$.  We define a second length scale,
\begin{equation}
r \equiv R/\Gamma. \label{rR}
\end{equation}
For simplicity, we ignore cosmological redshift.  In the frame of an
external observer, the time since the explosion is $\sim R/c$, while
in the comoving frame of the expanding shell, the time is $\sim r/c$.
The causal horizon around any point in the fluid is thus a sphere of
radius $\sim r$.  We assume causal contact in the radial direction.
The radial width of the shell in the fluid frame is then $\sim r$, and
the radial width in the observer frame is $\sim r/\Gamma =
R/\Gamma^2$.

The radiation from any fluid element in the shell will be isotropic in
the fluid frame, but beamed within a cone of half-angle $\sim1/\Gamma$
in the observer frame.  For a given distant observer, most of the
received radiation comes from a circular patch of transverse radius
$\sim R/\Gamma= r$ on the shell, i.e., from a single causal volume in
the radiating fluid.

Consider the radiation emitted from the outer surface of the shell.
As seen by the observer, the time delay between the radiation received
from the center of the visible patch and that from the edge of the
patch is $\sim R/\Gamma^2 c$ (we ignore factors of order unity in this
paper).  Assuming the shell is homogeneous, this is the shortest time
scale over which the observed signal can vary.  Since the radial width
of the shell in the observer frame is $\sim R/\Gamma^2$, the smoothing
time due to the finite radial width of the source is also of the same
order.  Thus, in this model of a GRB, which we refer to as the {\it
standard model}, the variability time scale is given by
\begin{equation}
{\rm Standard~Model:}\qquad\qquad t_{\rm var} \sim R/\Gamma^2c.
\label{eqstd}
\end{equation}

It is natural to associate the time scale $t_{\rm var}$ with the
duration of individual pulses (spikes) in the $\gamma$-ray lightcurve.
However, for a typical long GRB, the total burst duration $t_{\rm
burst}$ is several tens of times, and sometimes even a couple of
hundred times, longer than $t_{\rm var}$.  To explain the longer time
scale $t_{\rm burst}$, the standard model invokes a long-lived central
engine with significant power output over a time $\sim t_{\rm burst}$.
Furthermore, the engine is postulated to be highly variable and to
eject a large number of successive shells with different Lorentz
factors.  These shells collide with one another in internal shocks,
each shock producing a pulse in the lightcurve of duration $t_{\rm
var}$.

Instead of having multiple shells and internal shocks, could the burst
variability be explained via inhomogeneity in the radiating fluid?
For instance, could $t_{\rm burst}$ be equal to $R/\Gamma^2 c$ and
could the observed rapid variations in the lightcurve be the result of
bursts of radiation from tiny active blobs within the radiating fluid?
Sari \& Piran (1997) gave the following simple and powerful argument
against this possibility.

Let us define the variability parameter ${\cal V} \equiv t_{\rm
burst}/t_{\rm var}$; a typical value is ${\cal V}\sim100$.  For most
bursts, ${\cal V}$ is roughly equal to the number of pulses in the
lightcurve, i.e., the pulses fill the lightcurve with a duty cycle of
order unity.  If we wish to set $t_{\rm burst}$ equal to
$R/\Gamma^2c$, then a blob that produces any single pulse in the
lightcurve must have a radial extent no larger than $\sim r/{\cal V}$.
Assuming that the blobs are roughly spherical in shape (in the
comoving frame of the fluid), this means that there must be $\sim
{\cal V}^3$ independent blobs within a causal volume ($\sim r^3$) of
the fluid.  However, the number of pulses observed in the GRB
lightcurve is no more than ${\cal V}$.  Also, each pulse must be
produced on average by only one blob since the intensity varies by
order unity across a pulse.  We thus conclude that, out of $\sim{\cal
V}^3$ blobs, only ${\cal V}$ blobs radiate\footnote{Sari \& Piran
(1997) considered a somewhat different geometry where they took the
radial width of blobs to be same as the shell width, and thus
concluded that ${\cal V}$ out of a total of ${\cal V}^2$ blobs
radiate, or that the radiative efficiency is $\sim {\cal
V}^{-1}$.}. That is, only one out of every ${\cal V}^2\sim10^4$ blobs
radiates, and $\sim99.99\%$ of the fluid is silent.

It is highly unlikely that the GRB energy is localized inside just
$\sim10^{-4}$ of the volume of the fluid in the shell.  It is more
likely that the energy from the explosion is spread uniformly over the
entire shell.  But if this is the case, then the prompt GRB emission
must be highly inefficient, with only $\sim10^{-4}$ of the available
energy being radiated during the GRB.  Such extreme inefficiency is
unpalatable.  For instance, after correcting for beaming, the energy
release in gamma-rays in a typical long-duration GRB is found to be of
order 10$^{51}$erg (Frail et al. 2001).  With an inhomogeneous model
in which the efficiency is only $\sim0.01\%$, the true energy release
would be $\sim10^{55}$~erg, which is larger by a factor $\sim10^4$
than the kinetic energy of relativistic ejecta in GRBs as determined
from multiwavelength modeling of their afterglow lightcurves (Wijers
\& Galama, 1999; Panaitescu \& Kumar, 2002).

We are thus compelled to give up the idea of variability arising from
inhomogeneity, and forced to accept the standard internal shock model.
According to this model, the burst duration $t_{\rm burst}$ is equal
to the lifetime of the central engine, variability is produced by a
large number of random internal shocks among independent shells
ejected from the engine, and the variability time $t_{\rm var}$ is
given by equation (\ref{eqstd}).

\section{Turbulent Model}

We now describe an alternative model --- the {\it turbulent model} ---
in which we assume that the fluid in the GRB shell is relativistically
turbulent.  In the shell frame, let the typical Lorentz factor of an
energy-bearing eddy be $\gamma_t$.  As mentioned earlier, the lifetime
of the system in the shell frame is $\sim r/c$.  In the frame of an
eddy, this corresponds to a lifetime $\sim r/\gamma_t c$.  Therefore,
by causality, we expect the maximum size of an eddy in its own frame
to be $\sim r/\gamma_t$.  Let us make the reasonable assumption that
the energy-bearing eddies have roughly this size.  Thus, the size of
an eddy in its own frame is
\begin{equation}
r_e \sim r/\gamma_t \sim R/\Gamma\gamma_t.
\end{equation}
Each eddy has a volume $\sim r_e^3$, so we expect the total number of
eddies in a causal volume of the shell to be
\begin{equation}
n_e \sim (r/r_e)^3 \sim \gamma_t^3.
\end{equation}
In the shell frame, an eddy has a size $\sim r_e$ in a plane
perpendicular to its velocity vector, and a Lorentz-contracted size
$\sim r_e/\gamma_t$ parallel to its motion.

Eddies are not likely to travel along perfectly straight lines.
Rather, we expect their velocities to change on approximately the
causal time, which is $\sim r/c$ in the shell frame.  We also expect
eddies to dissolve and reform on this time scale.  However, since
$r/c$ is roughly the lifetime of the system we do not expect multiple
generations of eddies.

Consider now the radiation from an eddy as viewed in the shell frame.
At any instant, the radiation is beamed into a cone of half-angle
$1/\gamma_t$.  During the life of the eddy, the orientation of the
beam wanders by a few radians as a result of turbulent acceleration.
Thus each eddy illuminates a total solid angle $\sim 1/\gamma_t$ in
the shell frame in the course of its motion.  Boosting to the observer
frame, the illuminated solid angle from each eddy is
$\sim1/\Gamma^2\gamma_t$.  Summing over all $n_e$ eddies in a causal
volume, the total solid angle illuminated by all the eddies is $\sim
\gamma_t^2/\Gamma^2$.  All of this radiation is beamed within a solid
angle $\sim 1/\Gamma^2$.  Therefore, each observer receives radiation
from $\sim \gamma_t^2$ eddies.

An observer receives radiation from the entire collection of eddies
(inside one causal volume) over a time $\sim R/\Gamma^2c$.  In a major
departure from the standard model, let us associate this time with the
burst duration $t_{\rm burst}$.  The radiation received from a single
eddy then corresponds to an individual pulse in the GRB lightcurve.
To estimate the duration of a pulse, we note that the thickness of an
eddy in a direction parallel to its beamed radiation is $\sim r/
\gamma_t^2$ in the shell frame, or $\sim R/\Gamma^2\gamma_t^2$ in the
observer frame.  Thus, an observer receives radiation from a single
eddy within a time $\sim t_{\rm burst}/\gamma_t^2$.  As we showed
above, an observer receives on average $\sim\gamma_t^2$ pulses.  Thus,
in the turbulent model, we have the following results:
\begin{eqnarray}
\qquad\qquad &~& t_{\rm burst} \sim R/\Gamma^2c, \label{tburst}\\
{\rm Turbulent~Model}:\qquad\qquad &~& t_{\rm var} \sim R/\Gamma^2\gamma_t^2c,
\label{tvar} \\
\qquad\qquad &~& n_{\rm pulse} \sim \gamma_t^2, \label{npulse}
\end{eqnarray}
where $n_{\rm pulse}$ is the mean number of pulses in the burst.

\section{Sample Lightcurve}

Equations (\ref{tburst})--(\ref{npulse}) show that $n_{\rm
pulse}t_{\rm var}\sim t_{\rm burst}$.  This has two implications.
First, it means that pulses typically fill the entire duration of the
burst, i.e., the duty cycle of the pulses is of order unity, as
observed in GRBs.  Second, an observer receives radiation on average
from only one eddy at any given time.  Thus, we expect order unity
variations in the observed $\gamma$-ray flux, again consistent with
observations.  These features are illustrated in the sample lightcurve
shown in Fig. 1.

\begin{figure}
\includegraphics[width=6.5in]{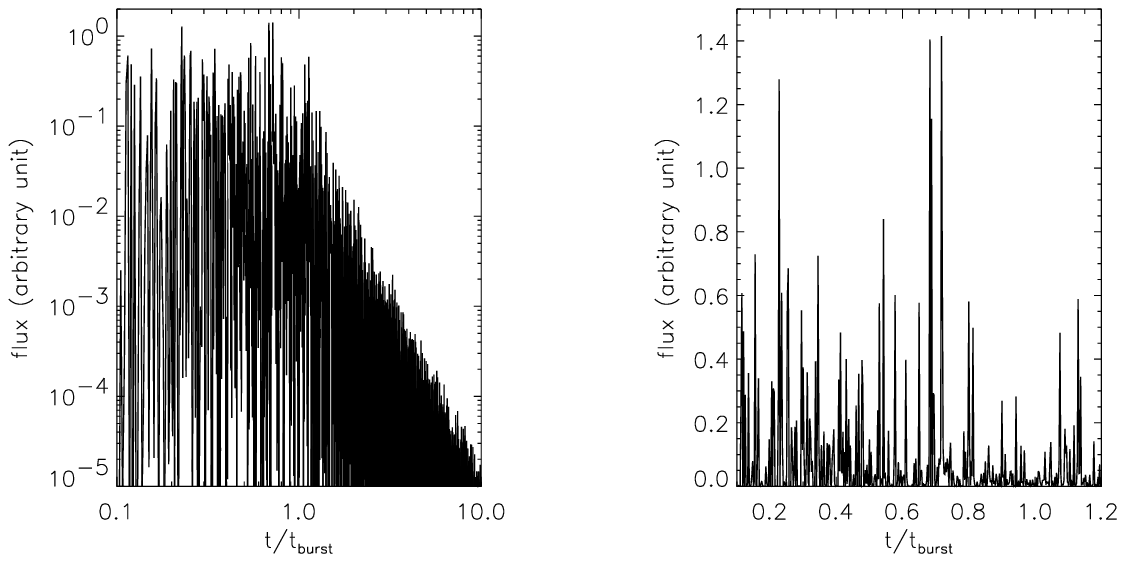}
\caption{Typical GRB lightcurve, calculated using the relativistic
turbulent model.  The two panels correspond to logarithmic and linear
scales, respectively.  The bulk Lorentz factor of the expanding GRB
shell is taken to be $\Gamma=500$ and the turbulent Lorentz factor of
the eddies to be $\gamma_t=10$.  Time is scaled by the burst duration
as defined in eq. (\ref{tburst}).  Note the high degree of variability
during the main burst ($t\ \lta\ t_{\rm burst}$), and the rapid
decrease of the flux at late times due to off-axis emission.}
\label{lightcurve} 
\end{figure}

Figure 1 was computed by considering a GRB shell expanding outward
with a large Lorentz factor $\Gamma=500$ (the precise value is
unimportant).  The shell is randomly filled with a population of
eddies with turbulent Lorentz factor $\gamma_t=10$.  It is assumed
that there are $\gamma_t^3$ eddies per volume $r^3$ in the frame of
the shell and that there is a probability $1/\gamma_t$ that the
relativistically boosted beam of any given eddy will sweep past the
observer during the eddy lifetime.  The spectral index of the
radiation is taken to be $\beta=1$ and the observed flux in a fixed
energy band is computed using standard relativistic transformations.

The resulting lightcurve depends to some extent on the precise
assumptions we make.  However, the model assumptions described above
are reasonable.  As Fig. 1 shows, this model gives a large number
($\sim\gamma_t^2$) of pulses during the main burst\footnote{In the
spirit of this paper, we have ignored numerical factors of order unity
in our definition of $t_{\rm burst}$ (eq. \ref{tburst}).  The burst
duration is probably closer to $R^2/2\Gamma^2c$.}, with a duty
cycle not very different from unity.  The rapidly declining late-time
flux is the result of off-axis emission.  Note that this emission
continues to show some residual variability.  This feature of the
model may be worth verifying through observations.  However, given the
rapid decline in the flux and the limited sensitivity of detectors,
one is generally forced to time-average the late-time data and this
will cause the variability to be washed out.

\section{Discussion}

The turbulent model described in this paper has several attractive
features.

\smallskip\noindent (i) The model naturally produces a highly variable
GRB lightcurve with large amplitude variations across individual
pulses and a duty cycle of order unity (\S4, Fig. 1).

\smallskip\noindent (ii) According to equation (\ref{tburst}), the
quantity $R/\Gamma^2c$ is equal to $t_{\rm burst}$ rather than $t_{\rm
var}$ (eq. \ref{eqstd}).  Relative to the standard model, the
turbulent model can thus accommodate much larger values of $R$ and
smaller values of $\Gamma$.  This eliminates a problematic constraint
which leads to difficulties when attempting to fit GRB observations
using equation (\ref{eqstd}) from the standard model (e.g., Kumar \&
Narayan 2008).

\smallskip\noindent (iii) The larger value of $R$ obtained with the
turbulent model is compatible with ideas described in Lyutikov \&
Blandford (2003), according to which the jet energy is primarily in
the form of Poynting flux.  This energy is converted to radiation
through plasma instabilities near the deceleration radius ($\lta\
10^{17}$ cm).  The same instabilities may also produce the
relativistic turbulence invoked in our model.

\smallskip\noindent (iv) The turbulent model neatly avoids the
efficiency argument of Sari \& Piran (1997).  Any particular observer
receives radiation from only a fraction $\sim1/\gamma_t$ of the
available eddies.  However, the additional relativistic boost of the
received radiation because of turbulent motion makes up for the
missing eddies.  Thus, each observer receives a fair share of the
emission from the shell, and there is no radiative inefficiency in the
model.

\smallskip\noindent (v) The model has a clear prediction for the
variability parameter ${\cal V}$:
\begin{equation}
{\cal V} \sim \gamma_t^2, \qquad \gamma_t \sim {\cal V}^{1/2}.
\end{equation}
Since a typical long GRB has ${\cal V}\sim100$, the turbulent eddy
Lorentz factor $\gamma_t$ needs to be $\sim 10$.

For simplicity, we have assumed in this paper that the eddy motions
are isotropic in the frame of the GRB shell.  This is, however, not
essential.  We could, for instance, have random relativistic motions
which are concentrated primarily in a plane perpendicular to the
radius vector of the shell, e.g., parallel to the local tangential
magnetic field.  Such a model would give qualitatively similar
results, though some of the scalings may be a little different.  Also,
we have simplified matters by assuming that all eddies have the same
Lorentz factor, which is unlikely in a real turbulent medium.  In
fact, a likely scenario is that a part of the fluid moves
relativistically with a range of Lorentz factors, and a part resides
in a (more-or-less) stationary inter-eddy medium which is produced
when eddies collide with one another in shocks.  This would slightly
modify the radiative properties of the medium (Kumar \& Narayan 2008),
but it will not change the key features of the model as described in
the present paper.  Finally, the relevant energy-bearing eddies may
not be as large as their comoving causality size but may be smaller by
a numerical factor, and the solid angle swept by an eddy in the shell
frame may be different from $\sim1/\gamma_t$ (for instance, eddies
might be hardly accelerated at all once they are formed).  These
effects will modify equations (\ref{tburst}), (\ref{tvar}) and
(\ref{npulse}).  However, the results will remain qualitatively the
same.

\acknowledgements

\end{document}